\documentclass[twocolumn,showpacs,preprintnumbers,amsmath,amssymb]{revtex4}

\usepackage{graphicx}
\usepackage{dcolumn}
\usepackage{bm}

\begin{document}

\title{Direct Observation of Sub-Poissonian Number Statistics in a Degenerate Bose Gas}

\author{C.-S. Chuu}
\author{F. Schreck}
 \altaffiliation[Present address: ]{Institut f\"{u}r
Quantenoptik und Quanteninformation, Innsbruck, Austria.}
\author{T. P. Meyrath}
\author{J. L. Hanssen}
 \altaffiliation[Present address: ]{National Institute of
Standards and Technology, Gaithersburg, MD 20899, USA.}
\author{G. N. Price}
\author{M. G. Raizen}
\affiliation{%
Center for Nonlinear Dynamics and Department of Physics,\\
The University of Texas at Austin, Austin, TX 78712, USA}%

\date{\today}

\begin{abstract}
We report the direct observation of sub-Poissonian number
fluctuation for a degenerate Bose gas confined in an optical trap.
Reduction of number fluctuations below the Poissonian limit is
observed for average numbers that range from 300 to 60 atoms.
\end{abstract}

\maketitle

The study of the quantum statistics of light has been at the heart
of modern quantum optics for many years, with examples ranging from
photon anti-bunching \cite{1} and squeezed states of light
\cite{2-5}, to quantum communication \cite{6}. The emerging field of
atom optics has now reached the stage where the direct measurement
of atom statistics can have a similar impact. In particular, novel
quantum statistics have been predicted for quantum degenerate gases
under a wide range of conditions. Two recent examples are the
prediction of sub-Poissonian statistics and Fock state production in
the Mott insulator transition \cite{7} and in the Quantum Tweezer
\cite{8}, and atomic anti-bunching in a one-dimensional gas of
repulsive bosons (Tonks-Girardeau gas) \cite{9}. More generally, it
is clear that the controlled study of entanglement and quantum
computing with massive particles must be based on the detection at
single-atom or ion level.

Following this theoretical work, an early experiment reported number
squeezing in an optical lattice based on the observation of
increased phase noise \cite{10}. Subsequent experiments have
provided clear and convincing evidence of the Mott insulator state
\cite{11-13}. In parallel work, several groups have observed novel
behavior of a 1-D gas in the Tonks regime \cite{14-15}. However all
of these experiments were conducted with a large number of atoms and
were therefore not statistical in nature.

In order to directly probe the atom statistics of these novel
states, one must incorporate single-atom counting with a
Bose-Einstein condensate (BEC) apparatus. We report in this Letter
the experimental realization of such a system and the first direct
measurement of sub-Poissonian atom number statistics in a degenerate
Bose gas.

The fluctuations of atom number within a small volume in a classical
ideal gas is given by $\sigma_{\it N}={\it N}^{1/2}$, where {\it N}
is the mean atom number \cite{Landau}. For the fluctuations of atom
number in a degenerate Bose gas this is not necessarily true and has
been the topic of intense theoretical debate. In the case of an
ideal Bose gas, number fluctuations have been studied in a box
\cite{18-20} and, more recently, in a harmonic trap
\cite{21-23,24-26} for microcanonical and canonical ensembles. For a
weakly interacting Bose gas, number fluctuations were first
investigated in Ref. \cite{27}. The role of interactions was then
further studied by including the effect of the thermal excitation of
phonons in the thermodynamic limit with number-nonconserving
\cite{28-1} and number-conserving \cite{28-2} Bogoliubov methods.
Most recently, an isolated system of finite atom number was
considered for studying number fluctuations in a harmonic trap
\cite{29} and in a box \cite{30}. The result obtained in the latter
case, in particular, predicted number fluctuation proportional to
$N^{1/2}$.

In BEC experiments reported thus far, typical shot-to-shot number
fluctuations greatly exceed the Poissonian limit, presumably due to
technical noise. Here we report on an ultra-stable optical trap
which has a controllable trapping volume and depth. This trap can be
used to achieve sub-Poissonian number fluctuations by the following
mechanism: for a fully loaded trap, the potential depth $U_0$ in the
shallowest direction is equal to the chemical potential $\mu$ of the
degenerate Bose gas if one neglects tunnelling out of the trap. In
the Thomas-Fermi (TF) limit, the atom number $N$ is proportional to
$\mu^{5/2}$ for a harmonic trapping potential. The atom number is
thus related to the trap depth as $N \propto U_0^{5/2}$. From this,
it is clear that a precise control of the trap depth can lead to a
precise control of the atom number. This conclusion remains valid
even when assuming tunnelling and a realistic nonharmonic potential.

\begin{figure}
\includegraphics{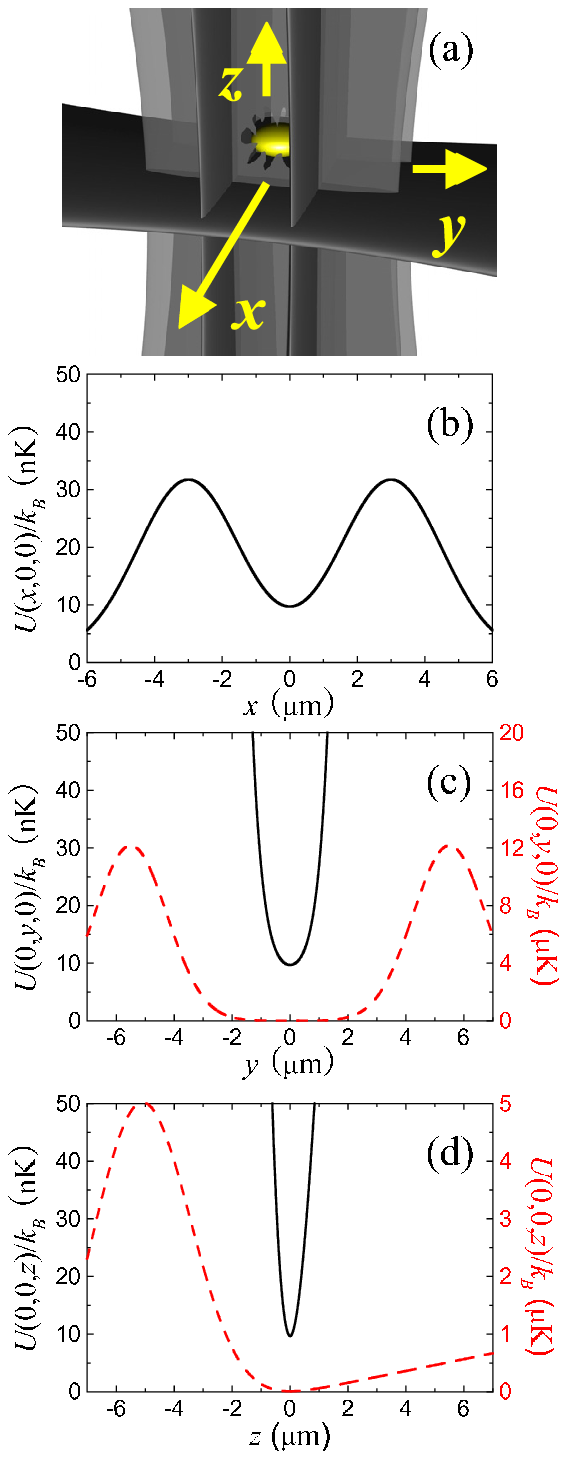}
\caption{\label{fig:1} (color online) The optical trap. (a) Beam
pictorial. Gravity points in -{\it z} direction in this pictorial.
(b) This plot shows the potential energy $U$ at $y=z=0$. In the $x$
direction, the atoms are confined by two Gaussian sheets with a
separation of $l_x=5\,{\rm \mu m}$. (c) Confinement in the $y$
direction is of similar shape to the $x$ direction but deeper. (d)
In the $z$ direction, gravity and a Gaussian sheet form a
gravito-optical trap. The dashed curves in (c) and (d) show the same
potentials for the different scales given on the right axes. Each
potential plot is calculated based on measured beam characteristics
and are appropriate for order 100 atoms.}
\end{figure}

Our experimental apparatus for studying sub-Poissonian number
statistics is similar to our previous work \cite{Our1}. A BEC of
$2\times10^5$ $^{87}$Rb atoms is produced by evaporation in a
large volume optical dipole trap. The BEC is then compressed and
transferred to the final small-volume optical trap. This trap is
formed by five Gaussian sheets, with two pairs propagating
vertically and one horizontal sheet to hold the atoms against
gravity, shown pictorially in Fig.~\ref{fig:1}(a). The calculated
potentials given by the measured beam parameters are shown in
Fig.~\ref{fig:1}(b) to Fig.~\ref{fig:1}(d) for the respective
directions. All beams originated from a 10\,W laser at $\lambda=$
532\,nm are tightly focused in one axis at the position of the
condensate. Each sheet pair is derived from the first order
deflections of multiple frequency acousto-optical modulators,
providing independent control of the position and power
\cite{Our2}. The sheet pairs and the horizontal sheet have a
maximum power of $P^{\rm max}_x=25$\,mW, $P^{\rm max}_y=80$\,mW,
$P^{\rm max}_z=100$\,mW per sheet and a $1/{\rm e}^2$ radius of
$2.5\,\mu$m $\times\; 100\,\mu$m for the {\it x} and {\it y} axes
and $3.4\,\mu$m $\times\; 200\,\mu$m for the {\it z} axis,
respectively; {\it x, y} and {\it z} refer to the potential axes
of Fig.~\ref{fig:1}. For typical operating conditions, the trap
has a depth of $U_0/k_B=$ 22\,nK (for $P_x=0.2\,$mW) with the
weakest trapping potential in the {\it x} direction and a
geometric mean trapping frequency of $\bar{\omega}=2\pi \times$
300\,Hz; $k_B$ is Boltzmann's constant. In the final evaporation
stage $U_0$ is ramped down adiabatically over a period of 1500\,ms
with an exponential shape. $U_0$, being the lowest evaporation
barrier, determines the chemical potential and thus the atom
number. Its final value is varied to obtain different atom
numbers.

Two methods are employed for measurements of atom numbers. For
numbers of order $10^3$ or larger, absorption imaging is used
yielding spatial and number information. At lower atom numbers
however, fluorescence imaging is used because of higher
signal-to-noise ratio in this regime. This is accomplished by
transferring the atoms into a small magneto-optical trap (MOT)
\cite{16-17}. The MOT uses six beams with a diameter of 1 mm, an
intensity of $65\,$mW/cm$^2$, a detuning of about $10\,$MHz, and a
magnetic field gradient of $260\,$G/cm. Transfer from the optical
trap to the MOT shows a saturation behavior with MOT beam intensity,
indicating that all atoms are captured. The resulting fluorescence
signal is detected by a charge-coupled-device (CCD) camera for
100\,ms and is calibrated against an avalanche photodiode (APD).
Because of the low density during exposure, there is little
possibility for multiple scattering events during detection.
Therefore, the measured fluorescence signal from the MOT is
proportional to the number of atoms present.

\begin{figure}
\includegraphics{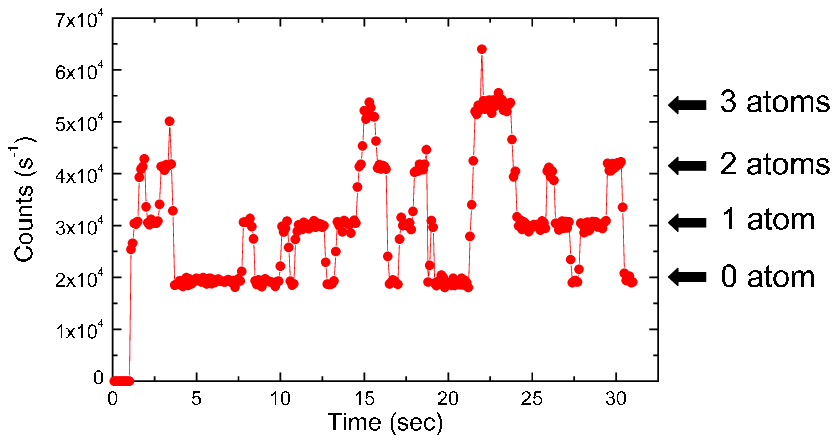}
\caption{\label{fig:2} (color online) Step-wise signal of the APD.
The atom number from CCD fluorescence imaging is calibrated by an
APD operating in photon counting mode. The fluorescence counting
rate per atom is $10^4\,{\rm s}^{-1}$ with a background of $2
\times 10^4\,{\rm s}^{-1}$. The signals shown are for random
loading of the MOT from background vapor. The time step of
fluorescence binning is 100\,ms.}
\end{figure}

The calibration of the atom number obtained from fluorescence
imaging is accomplished by operating the MOT in a regime
\cite{SAD} where discrete fluorescence levels of different small
atom numbers are observable on an APD, as shown in
Fig.~\ref{fig:2}. The MOT is then suddenly switched to the typical
operating settings described above and the new fluorescence level
is obtained for the same atom number. This yields an absolute
accuracy in atom number better than $\pm10\,\%$ \cite{abs}. The
result is consistent with calculations of the ratio of scattering
rates for the given settings \cite{Metcaft}. Atom numbers may also
be estimated by using the TF approximation with $\mu=U_0$ and
modelling the trap by a harmonic potential. This yields a $35\,\%$
deviation below the measured atom numbers, indicating rough
accuracy of this model.

Due to inhomogeneity of the optical potential, after the potential
barrier is lowered, some atoms remain outside of the trap. These
atoms are removed before the final number detection by raising the
potential barrier to its maximum intensity ($U^{\rm
max}_0/k_B=3\,\mu$K) and using a supplementary optical sheet pair
to sweep the residual atoms away from the well. A magnetic
gradient is also applied to remove atoms outside the range of the
sweeping beams.

\begin{figure}
\includegraphics{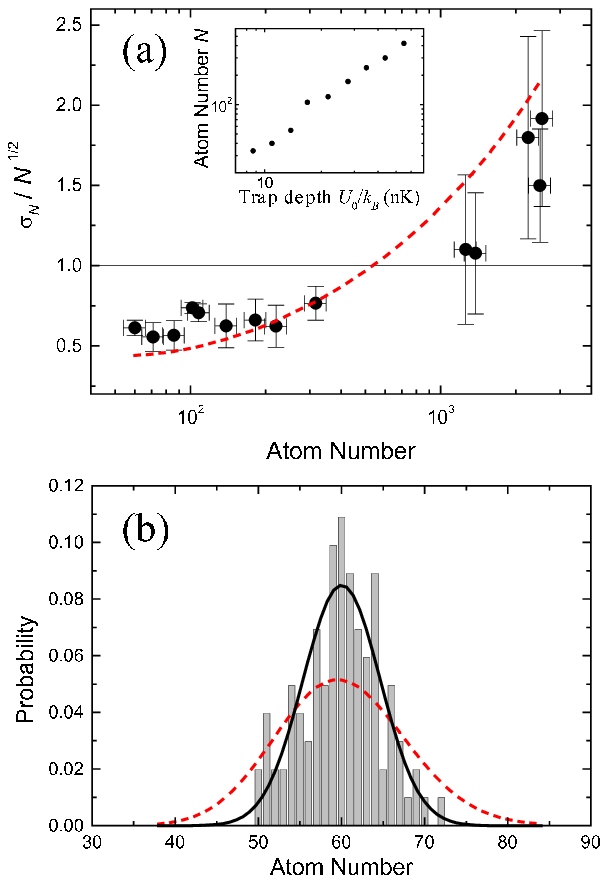}
\caption{\label{fig:3} (color online) Observation of sub-Poissonian
number statistics. (a) The solid circles show the measured
fluctuation normalized to the Poissonian case as a function of atom
number. The vertical error bars are the $68\,\%$ confidence
intervals for each measurement. The horizontal error bars represent
the absolute accuracy in atom number. Sub-Poissonian number
statistics is observed for atom numbers below 500 where the
fluctuations due to technical noise are not larger than the
Poissonian fluctuation. The dashed curve is the estimated
fluctuation from technical noise and background capture. The inset
shows the measured atom number as a function of the trap depth
$U_0$. (b) Histogram of 100 sample points shows sub-Poissonian
statistics for mean atom number $N=60$. The solid curve is the
Gaussian fit to the distribution. The dashed curve is the Poissonian
distribution with the same mean atom number.}
\end{figure}

Fig.~\ref{fig:3}(a) shows the measured fluctuations normalized to
the Poissonian case $\sigma_N/N^{1/2}$ (solid circles) as a function
of atom number. Sub-Poissonian fluctuations are observed for atom
numbers below 500, where the technical noise is no longer dominant
\cite{noise}. The measured fluctuation at $N=60$ atoms is
approximately $60\,\%$ of the corresponding Poissonian fluctuation.
This series of 100 measurements is shown as a histogram plot in
Fig.~\ref{fig:3}(b). A Gaussian fit to the data indicates a measured
standard deviation of $\sigma_N/N=7.9\,\%$ with a $99\,\%$
confidence interval of $[\,7.4\,\%,8.5\,\%\,]$ \cite{stat}. This
indicates a reliable measurement of deviation well below the
Poissonian value of $N^{1/2}/N=(12.9 \pm 0.5)\,\%$, where the error
is given by absolute accuracy in atom number. Several possible
technical noise sources were measured and estimates of their
contribution to the atom number fluctuation result in the following
\cite{noise}: $2.0\,\%$ from $P_x$, $2.4\,\%$ from $P_y$, $0.1\,\%$
from $P_z$, $2.2\,\%$ from $l_x$, and $2.0\,\%$ from $l_y$. The
overall contribution due to technical noise gives an expected atom
number fluctuation of $\delta_{\rm tech}=4.3\,\%$, which is very
close to what is measured for larger atom numbers. For lower atom
numbers, background capture during the detection is a major
contribution to the measured fluctuation. The background capture
during the 100 ms image-taking process with no atoms present is
measured to have a mean of $N_{\rm bg}=5$ atoms. This random
process, which has Poissonian statistics, broadens the measured atom
number distribution. A simple estimate of atom number fluctuation
from both technical noise and background capture, assuming a
constant background capture for different atom number $N$, results
in $\sqrt{(\delta_{\rm tech})^2+((N_{\rm bg})^{1/2}/N)^2}$, shown as
the dashed curve in Fig.~\ref{fig:3}(a). This result gives a similar
increasing tendency as for the measured fluctuations. A more
detailed calculation without TF approximation shows an increase in
sensitivity to trap fluctuations at lower atom numbers \cite{Artem}.

\begin{figure}
\includegraphics{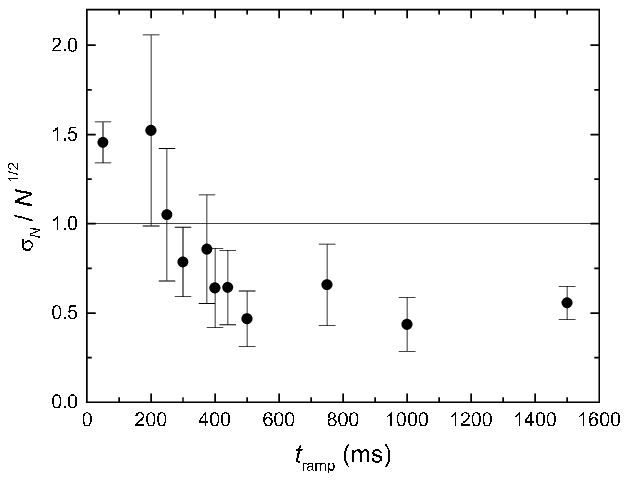}
\caption{\label{fig:4} Atom number fluctuations (normalized to the
Poissonian case) as a function of ramp time. Adiabaticity breaks
down when the ramp time is shorter than 250\,ms. The error bars
show the $68\,\%$ confidence intervals for each measurement. The
horizontal line shows the level of Poissonian fluctuation.}
\end{figure}

We have measured the dependence of atom statistics on the ramp time,
$t_{\rm ramp}$, in order to probe the many-body dynamics. The
results are displayed in Fig.~\ref{fig:4} where the fluctuations
(normalized to the Poissonian case) are plotted as a function of
$t_{\rm ramp}$. We find that for time scales shorter than 250 ms,
the atom statistics become super-Poissonian, while for longer times
they are sub-Poissonian. This result provides the timescale for
adiabatic following, a key feature of the process. The theoretical
analysis of our system is yet to be completed, and requires the
development of a time-dependent many-body theory without a
mean-field approximation. Surveying previous theoretical work, the
closest case we have found is the analysis of relative number
fluctuations between two condensates separated by a tunnel barrier
that is ramped up in time \cite{twomode}. The authors of that paper
found sub-Poissonian fluctuations in the relative atom number under
appropriate conditions. However our system is considerably
different, with a single trap in a quantum degenerate regime
undergoing loss of atoms as the barrier is lowered. Recent work on
quantum kinetic theory may provide insights to the present system
\cite{Gardiner1-3} and finite-size trap effects are also being
considered \cite{finitewell}.

In conclusion, we have observed sub-Poissonian number statistics
in a degenerate Bose gas prepared in an optical dipole trap. By
precisely controlling the chemical potential, we obtained a wide
range of atom numbers starting at several tens and going up to a
few thousand atoms. For atom numbers below a few hundred, we
observed sub-Poissonian number statistics. Future work will be to
eliminate sources of technical noise and to approach the $N=1$
Fock state.

The authors would like to acknowledge support from the Sid W.
Richardson Foundation, the National Science Foundation, the R. A.
Welch Foundation, and the Alexander von Humboldt Foundation. We also
thank A. M. Dudarev and Q. Niu for useful discussions.

\bibliography{squeezing}

\begin{thebibliography}{30}
\expandafter\ifx\csname natexlab\endcsname\relax\def\natexlab#1{#1}\fi
\expandafter\ifx\csname bibnamefont\endcsname\relax
  \def\bibnamefont#1{#1}\fi
\expandafter\ifx\csname bibfnamefont\endcsname\relax
  \def\bibfnamefont#1{#1}\fi
\expandafter\ifx\csname citenamefont\endcsname\relax
  \def\citenamefont#1{#1}\fi
\expandafter\ifx\csname url\endcsname\relax
  \def\url#1{\texttt{#1}}\fi
\expandafter\ifx\csname urlprefix\endcsname\relax\def\urlprefix{URL }\fi
\providecommand{\bibinfo}[2]{#2}
\providecommand{\eprint}[2][]{\url{#2}}

\bibitem[{\citenamefont{Kimble et~al.}(1977)\citenamefont{Kimble, Dagenais, and
  Mandel}}]{1}
\bibinfo{author}{\bibfnamefont{H.~J.} \bibnamefont{Kimble}},
  \bibinfo{author}{\bibfnamefont{M.}~\bibnamefont{Dagenais}}, \bibnamefont{and}
  \bibinfo{author}{\bibfnamefont{L.}~\bibnamefont{Mandel}},
  \bibinfo{journal}{Phys.\ Rev.\ Lett.} \textbf{\bibinfo{volume}{39}},
  \bibinfo{pages}{691} (\bibinfo{year}{1977}).

\bibitem[{2-5()}]{2-5}
\bibinfo{note}{R. E. Slusher, L. W. Hollberg, B. Yurke, J. C. Mertz, and J. F.
  Valley, Phys. Rev. Lett. {\bf 55}, 2409 (1985); L.-A. Wu, H. J. Kimble, J. L.
  Hall, and H. Wu, Phys. Rev. Lett. {\bf 57}, 2520 (1986); M. G. Raizen, L. A.
  Orozco, M. Xiao, T. L. Boyd, and H. J. Kimble, Phys. Rev. Lett. {\bf 59}, 198
  (1987); Y. Yamamoto and H. A. Haus, Rev. Mod. Phys. {\bf 58}, 1001 (1986)}.

\bibitem[{\citenamefont{Braunstein and van Loock}(2005)}]{6}
\bibinfo{author}{\bibfnamefont{S.~L.} \bibnamefont{Braunstein}}
  \bibnamefont{and} \bibinfo{author}{\bibfnamefont{P.}~\bibnamefont{van
  Loock}}, \bibinfo{journal}{Rev.\ Mod.\ Phys.} \textbf{\bibinfo{volume}{77}},
  \bibinfo{pages}{513} (\bibinfo{year}{2005}).

\bibitem[{\citenamefont{Jaksch et~al.}(1998)\citenamefont{Jaksch, Bruder,
  Cirac, Gardiner, and Zoller}}]{7}
\bibinfo{author}{\bibfnamefont{D.}~\bibnamefont{Jaksch}},
  \bibinfo{author}{\bibfnamefont{C.}~\bibnamefont{Bruder}},
  \bibinfo{author}{\bibfnamefont{J.~I.} \bibnamefont{Cirac}},
  \bibinfo{author}{\bibfnamefont{C.~W.} \bibnamefont{Gardiner}},
  \bibnamefont{and} \bibinfo{author}{\bibfnamefont{P.}~\bibnamefont{Zoller}},
  \bibinfo{journal}{Phys.\ Rev.\ Lett.} \textbf{\bibinfo{volume}{81}},
  \bibinfo{pages}{3108} (\bibinfo{year}{1998}).

\bibitem[{\citenamefont{Diener et~al.}(2002)\citenamefont{Diener, Wu, Raizen,
  and Niu}}]{8}
\bibinfo{author}{\bibfnamefont{R.~B.} \bibnamefont{Diener}},
  \bibinfo{author}{\bibfnamefont{B.}~\bibnamefont{Wu}},
  \bibinfo{author}{\bibfnamefont{M.~G.} \bibnamefont{Raizen}},
  \bibnamefont{and} \bibinfo{author}{\bibfnamefont{Q.}~\bibnamefont{Niu}},
  \bibinfo{journal}{Phys.\ Rev.\ Lett.} \textbf{\bibinfo{volume}{89}},
  \bibinfo{pages}{070401} (\bibinfo{year}{2002}).

\bibitem[{\citenamefont{Kheruntsyan et~al.}(2003)\citenamefont{Kheruntsyan,
  Gangardt, Drummond, and Shlyapnikov}}]{9}
\bibinfo{author}{\bibfnamefont{K.~V.} \bibnamefont{Kheruntsyan}},
  \bibinfo{author}{\bibfnamefont{D.~M.} \bibnamefont{Gangardt}},
  \bibinfo{author}{\bibfnamefont{P.~D.} \bibnamefont{Drummond}},
  \bibnamefont{and} \bibinfo{author}{\bibfnamefont{G.~V.}
  \bibnamefont{Shlyapnikov}}, \bibinfo{journal}{Phys.\ Rev.\ Lett.}
  \textbf{\bibinfo{volume}{91}}, \bibinfo{pages}{040403}
  (\bibinfo{year}{2003}).

\bibitem[{\citenamefont{Orzel et~al.}(2001)\citenamefont{Orzel, Tuchman,
  Fenselau, Yasuda, and Kasevich}}]{10}
\bibinfo{author}{\bibfnamefont{C.}~\bibnamefont{Orzel}},
  \bibinfo{author}{\bibfnamefont{A.~K.} \bibnamefont{Tuchman}},
  \bibinfo{author}{\bibfnamefont{M.~L.} \bibnamefont{Fenselau}},
  \bibinfo{author}{\bibfnamefont{M.}~\bibnamefont{Yasuda}}, \bibnamefont{and}
  \bibinfo{author}{\bibfnamefont{M.~A.} \bibnamefont{Kasevich}},
  \bibinfo{journal}{Science} \textbf{\bibinfo{volume}{291}},
  \bibinfo{pages}{2386} (\bibinfo{year}{2001}).

\bibitem[{11-()}]{11-13}
\bibinfo{note}{M. Greiner, O. Mandel, T. Esslinger, T. W. H{\"{a}}nsch, and I.
  Bloch, Nature {\bf 415}, 39 (2002); M. Greiner, O. Mandel, T. W.
  H{\"{a}}nsch, and I. Bloch, Nature {\bf 419}, 51 (2002); F. Gerbier, A.
  Widera, S. F{\"{o}}lling, O. Mandel, T. Gericke, and I. Bloch, Phys. Rev.
  Lett. {\bf 95}, 050404 (2005)}.

\bibitem[{14-()}]{14-15}
\bibinfo{note}{T. Kinoshita, T. Wenger, and D. S. Weiss, Science {\bf 305},
  1125 (2004); B. Paredes, A. Widera, V. Murg, O. Mandel, S. F{\"{o}}lling, I.
  Cirac, G. V. Shlyapnikov, T. W. H{\"{a}}nsch, and I. Bloch, Nature {\bf 429},
  277 (2004)}.

\bibitem[{\citenamefont{Landau and Lifshitz}(1998)}]{Landau}
\bibinfo{author}{\bibfnamefont{L.~D.} \bibnamefont{Landau}} \bibnamefont{and}
  \bibinfo{author}{\bibfnamefont{E.~M.} \bibnamefont{Lifshitz}},
  \emph{\bibinfo{title}{Statistical Physics}}
  (\bibinfo{publisher}{Butterworth-Heinemann}, \bibinfo{address}{Oxford},
  \bibinfo{year}{1998}), \bibinfo{edition}{3rd} ed.

\bibitem[{18-()}]{18-20}
\bibinfo{note}{E. H. Hauge, Physica Nor. {\bf 4}, 19 (1969); I. Fujiwara, D.
  ter Haar, and H. Wergeland, J. Stat. Phys. {\bf 2}, 329 (1970); R. M. Ziff,
  G. E. Uhlenbeck, and M. Kac, Phys. Rep. {\bf 32}, 169 (1977)}.

\bibitem[{21-()}]{21-23}
\bibinfo{note}{H. D. Politzer, Phys. Rev. A {\bf 54}, 5048 (1996); M. Gajda and
  K. Rz{\c{a}}{\.{z}}ewski, Phys. Rev. Lett. {\bf 78}, 2686 (1997); P. Navez,
  D. Bitouk, M. Gajda, Z. Idziaszek, and K. Rz{\c{a}}{\.{z}}ewski, Phys. Rev.
  Lett. {\bf 79}, 1789 (1997)}.

\bibitem[{24-()}]{24-26}
\bibinfo{note}{M. Wilkens and C. Weiss, J. Mod. Opt. {\bf 44}, 1801 (1997); S.
  Grossmann and M. Holthaus, Phys. Rev. Lett. {\bf 79}, 3557 (1997); N. L.
  Balazs and T. Bergeman, Phys. Rev. A {\bf 58}, 2359 (1998)}.

\bibitem[{\citenamefont{Buffet and Pul{\`{e}}}(1983)}]{27}
\bibinfo{author}{\bibfnamefont{E.}~\bibnamefont{Buffet}} \bibnamefont{and}
  \bibinfo{author}{\bibfnamefont{J.~V.} \bibnamefont{Pul{\`{e}}}},
  \bibinfo{journal}{J.\ Math.\ Phys.} \textbf{\bibinfo{volume}{24}},
  \bibinfo{pages}{1608} (\bibinfo{year}{1983}).

\bibitem[{\citenamefont{Giorgini et~al.}(1998)\citenamefont{Giorgini,
  Pitaevskii, and Stringari}}]{28-1}
\bibinfo{author}{\bibfnamefont{S.}~\bibnamefont{Giorgini}},
  \bibinfo{author}{\bibfnamefont{L.~P.} \bibnamefont{Pitaevskii}},
  \bibnamefont{and}
  \bibinfo{author}{\bibfnamefont{S.}~\bibnamefont{Stringari}},
  \bibinfo{journal}{Phy.\ Rev.\ Lett.} \textbf{\bibinfo{volume}{80}},
  \bibinfo{pages}{5040} (\bibinfo{year}{1998}).

\bibitem[{\citenamefont{Kocharovsky et~al.}(2000)\citenamefont{Kocharovsky,
  Kocharovsky, and Scully}}]{28-2}
\bibinfo{author}{\bibfnamefont{V.~V.} \bibnamefont{Kocharovsky}},
  \bibinfo{author}{\bibfnamefont{V.~V.} \bibnamefont{Kocharovsky}},
  \bibnamefont{and} \bibinfo{author}{\bibfnamefont{M.~O.}
  \bibnamefont{Scully}}, \bibinfo{journal}{Phy.\ Rev.\ Lett.}
  \textbf{\bibinfo{volume}{84}}, \bibinfo{pages}{2306} (\bibinfo{year}{2000}).

\bibitem[{\citenamefont{Liu et~al.}(2003)\citenamefont{Liu, Xiong, Huang, and
  Xu}}]{29}
\bibinfo{author}{\bibfnamefont{S.}~\bibnamefont{Liu}},
  \bibinfo{author}{\bibfnamefont{H.}~\bibnamefont{Xiong}},
  \bibinfo{author}{\bibfnamefont{G.}~\bibnamefont{Huang}}, \bibnamefont{and}
  \bibinfo{author}{\bibfnamefont{Z.}~\bibnamefont{Xu}},
  \bibinfo{journal}{Phys.\ Rev.\ A} \textbf{\bibinfo{volume}{68}},
  \bibinfo{pages}{065601} (\bibinfo{year}{2003}).

\bibitem[{\citenamefont{Idziaszek et~al.}(1999)\citenamefont{Idziaszek, Gajda,
  Navez, Wilkens, and Rz{\c{a}}{\.{z}}ewski}}]{30}
\bibinfo{author}{\bibfnamefont{Z.}~\bibnamefont{Idziaszek}},
  \bibinfo{author}{\bibfnamefont{M.}~\bibnamefont{Gajda}},
  \bibinfo{author}{\bibfnamefont{P.}~\bibnamefont{Navez}},
  \bibinfo{author}{\bibfnamefont{M.}~\bibnamefont{Wilkens}}, \bibnamefont{and}
  \bibinfo{author}{\bibfnamefont{K.}~\bibnamefont{Rz{\c{a}}{\.{z}}ewski}},
  \bibinfo{journal}{Phys.\ Rev.\ Lett.} \textbf{\bibinfo{volume}{82}},
  \bibinfo{pages}{4376} (\bibinfo{year}{1999}).

\bibitem[{\citenamefont{Meyrath
  et~al.}(2005{\natexlab{a}})\citenamefont{Meyrath, Schreck, Hanssen, Chuu, and
  Raizen}}]{Our1}
\bibinfo{author}{\bibfnamefont{T.~P.} \bibnamefont{Meyrath}},
  \bibinfo{author}{\bibfnamefont{F.}~\bibnamefont{Schreck}},
  \bibinfo{author}{\bibfnamefont{J.~L.} \bibnamefont{Hanssen}},
  \bibinfo{author}{\bibfnamefont{C.-S.} \bibnamefont{Chuu}}, \bibnamefont{and}
  \bibinfo{author}{\bibfnamefont{M.~G.} \bibnamefont{Raizen}},
  \bibinfo{journal}{Opt.\ Express} \textbf{\bibinfo{volume}{13}},
  \bibinfo{pages}{2843} (\bibinfo{year}{2005}{\natexlab{a}}).

\bibitem[{\citenamefont{Meyrath
  et~al.}(2005{\natexlab{b}})\citenamefont{Meyrath, Schreck, Hanssen, Chuu, and
  Raizen}}]{Our2}
\bibinfo{author}{\bibfnamefont{T.~P.} \bibnamefont{Meyrath}},
  \bibinfo{author}{\bibfnamefont{F.}~\bibnamefont{Schreck}},
  \bibinfo{author}{\bibfnamefont{J.~L.} \bibnamefont{Hanssen}},
  \bibinfo{author}{\bibfnamefont{C.-S.} \bibnamefont{Chuu}}, \bibnamefont{and}
  \bibinfo{author}{\bibfnamefont{M.~G.} \bibnamefont{Raizen}},
  \bibinfo{journal}{Phys.\ Rev.\ A} \textbf{\bibinfo{volume}{71}},
  \bibinfo{pages}{041604(R)} (\bibinfo{year}{2005}{\natexlab{b}}).

\bibitem[{16-()}]{16-17}
\bibinfo{note}{Z. Hu and H. J. Kimble, Opt. Lett. {\bf 19}, 1888 (1994); S.
  Kuhr, W. Alt, D. Schrader, M. M{\"{u}}ller, V. Gomer, and D. Meschede,
  Science {\bf 293}, 278 (2001)}.

\bibitem[{SAD()}]{SAD}
\bibinfo{note}{A MOT intensity of $8$\,mW/cm$^2$ and a detuning of about
  $5$\,MHz are used for this measurement}.

\bibitem[{abs()}]{abs}
\bibinfo{note}{For absorption imaging, the absolute accuracy in atom number is
  estimated to be $\pm25\,\%$}.

\bibitem[{\citenamefont{Metcalf and van~der Straten}(1999)}]{Metcaft}
\bibinfo{author}{\bibfnamefont{H.~J.} \bibnamefont{Metcalf}} \bibnamefont{and}
  \bibinfo{author}{\bibfnamefont{P.}~\bibnamefont{van~der Straten}},
  \emph{\bibinfo{title}{Laser Cooling and Trapping}}
  (\bibinfo{publisher}{Springer}, \bibinfo{address}{New York},
  \bibinfo{year}{1999}).

\bibitem[{noi()}]{noise}
\bibinfo{note}{The rms technical noise is measured to be less than or equal to
  the following: $\Delta P_x=1\,\%$, $\Delta P_y=5\,\%$, $\Delta P_z=1\,\%$,
  $\Delta l_x=0.5\,\%$, and $\Delta l_y=0.2\,\%$. Based on TF approximation in
  the three-dimensional regime with harmonic potentials, an estimate of the
  atom number fluctuation caused by technical noise gives $4.3\,\%$. This
  limits the observation of sub-Poissonian fluctuations to $N<540$, as from
  $4.3\,\%<N^{-1/2}$}.

\bibitem[{sta()}]{stat}
\bibinfo{note}{A 99 $\%$ {\it confidence interval} is defined as
  $[\,\sqrt{\frac{n-1}{\chi^2_{0.005}(n-1)}}\sigma_N,
  \sqrt{\frac{n-1}{\chi^2_{0.995}(n-1)}}\sigma_N\,]$, where $\sigma_N$ is the
  measurement standard deviation, {\it n} is the number of samples, and
  $\chi^2_{0.005}$({\it n}-1), $\chi^2_{0.995}$({\it n}-1) are the upper
  $0.5\,\%$, $99.5\,\%$ points of the chi-square distribution, respectively,
  with $n-1$ degrees of freedom. See, for example, R. V. Hogg and E. A. Tanis,
  {\it Probability and Statistical Inference} (Prentice Hall, New Jersey,
  1997), 5th ed}.

\bibitem[{Art()}]{Artem}
\bibinfo{note}{A. M. Dudarev, M. G. Raizen, and Q. Niu, in preparation}.

\bibitem[{\citenamefont{Javanainen and Ivanov}(1999)}]{twomode}
\bibinfo{author}{\bibfnamefont{J.}~\bibnamefont{Javanainen}} \bibnamefont{and}
  \bibinfo{author}{\bibfnamefont{M.~Y.} \bibnamefont{Ivanov}},
  \bibinfo{journal}{Phys.\ Rev.\ A} \textbf{\bibinfo{volume}{60}},
  \bibinfo{pages}{2351} (\bibinfo{year}{1999}).

\bibitem[{Gar()}]{Gardiner1-3}
\bibinfo{note}{C.W. Gardiner, P. Zoller, R. J. Ballagh, and M. J. Davis, Phys.
  Rev. Lett. {\bf 79}, 1793 (1997); C.W. Gardiner, M.D. Lee, R. J. Ballagh, M.
  J. Davis, and P. Zoller, Phys. Rev. Lett. {\bf 81}, 5266 (1998); M. D. Lee
  and C.W. Gardiner, Phys. Rev. A {\bf 62}, 033606 (2000)}.

\bibitem[{\citenamefont{Carr et~al.}(2005)\citenamefont{Carr, Holland, and
  Malomed}}]{finitewell}
\bibinfo{author}{\bibfnamefont{L.~D.} \bibnamefont{Carr}},
  \bibinfo{author}{\bibfnamefont{M.~J.} \bibnamefont{Holland}},
  \bibnamefont{and} \bibinfo{author}{\bibfnamefont{B.~A.}
  \bibnamefont{Malomed}}, \bibinfo{journal}{J.\ Phys.\ B}
  \textbf{\bibinfo{volume}{38}}, \bibinfo{pages}{3217} (\bibinfo{year}{2005}).

\end{thebibliography}

\end{document}